\documentclass[a4paper,fleqn,usenatbib]{mnras}

\usepackage{newtxtext,newtxmath}

\usepackage[T1]{fontenc}
\usepackage{ae,aecompl}

\usepackage{graphicx}	
\usepackage{natbib,times}
\usepackage{array}

\title[`On the Red Supergiant Problem': a rebuttal]{On `On the Red Supergiant Problem': a rebuttal, and a consensus on the upper mass cutoff for II-P progenitors}

\author[B. Davies \& E. Beasor]{
Ben Davies,$^{1}$\thanks{b.davies@ljmu.ac.uk} and Emma R.\ Beasor$^{2}\thanks{Hubble Fellow}$
\\
$^{1}$Astrophysics Research Institute, Liverpool John Moores 
University, Liverpool Science Park ic2, \\ 146 Brownlow Hill, Liverpool, L3 5RF, UK\\
$^{2}$NSF's National Optical-Infrared Astronomy Research Laboratory, 950 N. Cherry Ave., Tucson, AZ 85721, USA
}

\date{}

\pubyear{2020}

\begin{document}
\label{firstpage}
\pagerange{\pageref{firstpage}--\pageref{lastpage}}
\maketitle
      
\begin{abstract}
The `Red Supergiant Problem' describes the claim that the brightest Red Supergiant (RSG) progenitors to type II-P supernovae are significantly fainter than RSGs in the field. This mismatch has been interpreted by several authors as being a manifestation of the mass threshold for the production of black holes (BHs), such that stars with initial masses above a cutoff of $M_{\rm hi}=17$M$_\odot$ and below 25$M_\odot$ will die as RSGs, but with no visible SN explosion as the BH is formed. However, we have previously cautioned that this cutoff is more likely to be higher and has large uncertainties ($M_{\rm hi}=19^{+4}_{-2}M_{\odot}$), meaning that the statistical significance of the RSG Problem is less than $2\sigma$. Recently, Kochanek (2020) has claimed that our work is statistically flawed, and with their analysis has argued that the upper mass cutoff is as low as $M_{\rm hi} = 15.7 \pm 0.8M_\odot$, giving the RSG Problem a significance of $>10\sigma$. In this letter, we show that Kochanek's low cutoff is caused by a statistical misinterpretation, and the associated fit to the progenitor mass spectrum can be ruled out at the 99.6\% confidence level. Once this problem is remedied, Kochanek's best fit becomes $M_{\rm hi} =19^{+4}_{-2}M_{\odot}$, in excellent agreement with our work. Finally, we argue that, in the search for a RSG `vanishing' as it collapses directly to a BH, any such survey would have to operate for decades before the absence of any such detection became statistically significant. 
\end{abstract}


\def\ga{\mathrel{\hbox{\rlap{\hbox{\lower4pt\hbox{$\sim$}}}\hbox{$>$}}}}
\def\la{\mathrel{\hbox{\rlap{\hbox{\lower4pt\hbox{$\sim$}}}\hbox{$<$}}}}
\def\msunyr{$M$ \mbox{$_{\normalsize\odot}$}\rm{yr}$^{-1}$}
\def\msun{$M$\mbox{$_{\normalsize\odot}$}}
\def\zsun{$Z$\mbox{$_{\normalsize\odot}$}}
\def\rsun{$R$\mbox{$_{\normalsize\odot}$}}
\def\minit{$M_{\rm init}$}
\def\mmax{$M_{\rm max}$}
\def\lsun{$L$\mbox{$_{\normalsize\odot}$}}
\def\mdot{$\dot{M}$}
\def\mdotdj{$\dot{M}_{\rm dJ}$}
\def\lbol{$L$\mbox{$_{\rm bol}$}}
\def\logl{$\log(L/L_\odot)$}
\def\kms{\,km~s$^{-1}$}
\def\EW{$W_{\lambda}$}
\def\arcsec{$^{\prime \prime}$}
\def\arcmin{$^{\prime}$}
\def\teff{$T_{\rm eff}$}
\def\Teff{$T_{\rm eff}$}
\def\logg{$\log g$}
\def\logz{$\log Z$}
\def\logl{$\log (L/L_\odot)$}
\def\vdisp{$v_{\rm disp}$}
\def\bcv{{\it BC$_V$}}
\def\bci{{\it BC$_I$}}
\def\bck{{\it BC$_K$}}
\def\lmax{$L_{\rm max}$}
\def\um{$\mu$m}
\def\chisq{$\chi^{2}$}
\def\AV{$A_{V}$}
\def\hminus{H$^{-}$}
\def\Hminus{H$^{-}$}
\def\ebmv{$E(B-V)$}
\def\mdyn{$M_{\rm dyn}$}
\def\mphot{$M_{\rm phot}$}
\def\cnterm{[C/N]$_{\rm term}$}
\newcommand{\fig}[1]{Fig.\ \ref{#1}}
\newcommand{\Fig}[1]{Figure \ref{#1}}
\newcommand{\newtext}[1]{{#1}}
\newcommand{\newnewtext}[1]{{#1}}
\def\gammaL{$\Gamma_{L}$}
\def\mhi{$M_{\rm hi}$}
\def\mlo{$M_{\rm lo}$}
\def\lhi{$L_{\rm hi}$}
\def\llo{$L_{\rm lo}$}
\def\lfin{$L_{\rm fin}$}

\def\llofit{$\log(L_{\rm lo}/L_\odot) = 4.39^{+0.10}_{-0.16}$}
\def\lhifit{$\log(L_{\rm hi}/L_\odot) = 5.20^{+0.17}_{-0.11}$}

\def\mlofit{$M_{\rm lo} = 6.9^{+0.9}_{-0.9}$\msun}
\def\mhifit{$M_{\rm hi} = 20.3^{+6.1}_{-2.6}$\msun}



\section{Introduction} \label{sec:intro}
Perhaps one of the greatest breakthroughs in massive star research of recent years has been the ability to directly associate supernovae (SNe) with their progenitor stars via archival pre-explosion imaging. Specifically, hydrogen-rich `plateau' supernovae (classified as II-P) have been unequivocally linked to Red Supergiant (RSG) progenitors \citep{smartt04,smartt09}. Once the progenitor is identified, it is possible (by adopting a series of key assumptions) to estimate the star's luminosity at death \lfin, and ultimately an initial mass \minit, providing a fundamental test of stellar evolutionary theory.  

In the first attempt to analyse a sample of II-P progenitors, \citet[][hereafter S09]{smartt09} noted that the most luminous progenitor \citep[SN1999ev with $\log(L/L_\odot) = 5.1$, but see Introduction of][]{Davies-Beasor20} was substantially fainter than the brightest RSGs in the field \citep[a luminosity limit often referred to as the `Humphreys-Davidson (H-D) limit', now established to be at $\log(L/L_\odot) = 5.5$, ][]{Humphreys-Davidson79,DCB18,Davies-Beasor20}. The sample consisted of 9 detections, and 12 upper limits, and the authors quoted the significance of this discrepancy to be 2.4$\sigma$, though Fig.\ 6 in S09 suggests rather it is somewhat below $2\sigma$ ($<90$\%). This possible tension between the II-P luminosity distribution and that of field RSGs was termed the `Red Supergiant Problem'.  

\newtext{Despite the statistical significance of the RSG Problem being within 3$\sigma$}, several explanations for its existence have subsequently appeared in the literature. A popular hypothesis has been that the `missing' RSGs (i.e. those which die with luminosities between $5.1 < \log(L/L_\odot) <5.5$) collapse to form black-holes (BHs) with no observable SN, which resonates somewhat with the results of independent numerical work \citep{OConnor-Ott11,Horiuchi14,Ertl16,Mueller16,Sukhbold18}. Converting these terminal luminosities, as well as the luminosity of the H-D limit, into initial masses using e.g. the STARS evolution code adopted by S09, this suggests that stars with initial masses between 17-25\msun\ will still evolve to become RSGs, but rather than explode as II-P SNe will simply vanish with no explosion. This has prompted searches for `disappearing' RSGs in archival survey data of nearby galaxies \citep[e.g.][]{Kochanek08}, but as yet no convincing example has been found \citep{Reynolds15,Adams17}. 



A more mundane explanation for the RSG Problem is that, \newtext{with its low statistical significance, it is possible that no problem exists at all}. In \citet[][ hereafter DB18]{Davies-Beasor18} we reanalysed the mass distribution of II-P progenitors, with improved measurements of foreground extinction from \citet{Maund17} and more realistic bolometric corrections. We adopted a Monte-Carlo style analysis method to determine the posterior probabilities on the lower and upper mass limits (\mlo, \mhi) to the distribution when both were allowed to be free parameters. Despite a larger sample, due to more nearby II-P SNe in the intervening years since S09, we argued that the significance of the RSG Problem was still less than 2$\sigma$. In \citet[][ hereafter DB20]{Davies-Beasor20} we studied the luminosity distribution rather than the (model dependent) mass distribution, removed the assumption of a Salpeter initial mass function, and looked more closely at the expectation value (i.e. the observations of the H-D limit). Again, we found a significance below 2$\sigma$. Nominally, DB20 found an upper mass limit \mhi=$18^{+4}_{-2}$\msun, again assuming the \minit-\lfin\ relation from the STARS models used by S09, and by comparing different evolutionary models estimated a further systematic error on \mhi\ of $\pm$1.3\msun. 

Recently, the analysis in DB18 has been challenged by \citet[][ hereafter K20]{Kochanek20}. In that paper, K20 adopt a Bayesian analysis method, initially finding results which are consistent with DB18 and DB20, specifically \mhi=$19^{+4}_{-2}$\msun\ (STARS \minit-\lfin\ relation). However, in performing a series of Monte-Carlo simulations with mock data, K20 found an apparent correlation between the upper error bar on \mhi, in this case +4\msun, and the fitted value of \mhi\ (see Fig. 5 in K20). This correlation was interpreted by K20 as being evidence that individual Monte Carlo simulations, randomly scattered about an input value by the observational errors, may be corrected back to that input value. K20 proceeded to use this correlation to adjust their best-fit value of \mhi\ obtained from analysis of the real data, arriving at \mhi=$15.8\pm0.8$\msun. This then gives a statistical significance to the RSG Problem of $>10\sigma$.

In Section \ref{sec:koch} of this paper, we will first rebut the conclusions of K20. We will show that the correlation observed between \mhi\ and its upper error bar is explained by error propagation when fitting a steep power law to data with non-zero errors. Furthermore, we will show that the low \mhi\ quoted by K20 is refuted by a comparison to the data used to derive it. Finally, in Section \ref{sec:bh} we will assess the prospects for observing RSGs spontaneously collapse to BHs.

\section{A rebuttal to Kochanek (2020)} \label{sec:koch}
In Section 3 of K20, \newnewtext{a Monte-Carlo (MC) experiment is performed} to demonstrate the accuracy and precision of the analysis method. The experiment involves randomly generating a sample of 24 masses from a power-law distribution characterized by a Salpeter slope ($x=1.35$), and input upper and lower mass limits \mlo$_{\rm, in}$=8\msun\ and \mhi$_{\rm, in}$=18\msun. These mock progenitor masses are randomly allocated to real progenitor sites to determine what their pre-explosion photometry would be, such that the posterior probability distributions on their inferred masses could be calculated. K20's Bayesian analysis is performed on each randomly-generated sample to obtain a two-parameter fit (\mhi\ and \mlo\ are allowed to vary, $x$ is fixed) to that trial's progenitor mass distribution. The results of 500 such trials are plotted in K20's Fig.\ 3, \newtext{demonstrating} a small systematic bias of \newnewtext{the} method (i.e. the offset of the cloud of points from the input values), as well as the random errors on \mhi\ and \mlo\ (i.e. the distribution of points about the median output values). 

\begin{figure}
\begin{center}
\includegraphics[width=8.5cm]{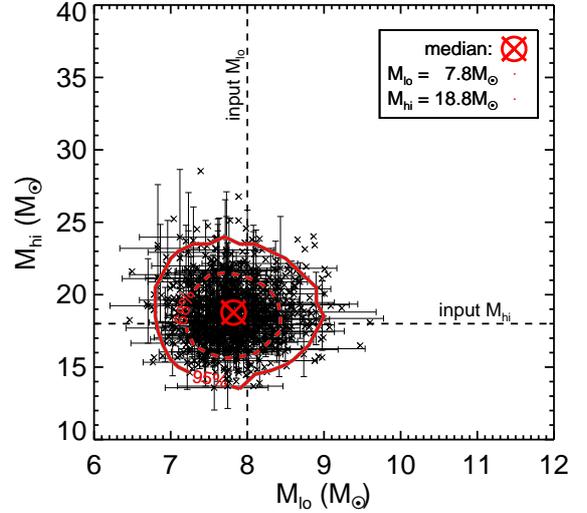}
\caption{Results of our Monte-Carlo experiment, analogous to K20's Fig.\ 3, showing the posterior distributions of \mlo\ and \mhi. As in K20, error bars are shown for only 10\% of the points for the sake of clarity. The red cross shows the median output \mlo\ and \mhi, both of which have a small systematic bias as in K20. The dashed and solid lines show the 68\% and 95\% confidence limits respectively, analogous to 1 and 2$\sigma$ random experimental error bars.}
\label{fig:kochMC}
\end{center}
\end{figure}

In \fig{fig:kochMC} of this current paper, we plot the results of a similar analysis of our own. Following K20, we generate 1000 samples of 24 randomly generated masses between \mlo$_{\rm, in}$=8\msun\ and \mlo$_{\rm, in}$=18\msun\ according to a power-law distribution with $x=-1.35$. To simulate experimental errors, we assume uniform fractional errors $\sigma_M$ on all masses, initially fixed at 20\%, but the effect of varying which is studied later. For each mass $M_i$ we randomly sample from the normal distribution centred on $\log(M_i) \pm \log(\sigma_M)$. We then fit these masses with the function,

\begin{equation}
M_i^{-x} = ( M_{\rm lo}^{-x} - f_i M_{\rm lo}^{-x} + f_i M_{\rm hi}^{-x} )
\label{equ:mf}
\end{equation}

\noindent where $x=1.35$ is again the Salpeter slope of the initial mass function, and $f_i$ is the normalised ranking of the $i$th supernova out of the sample of 24, ordered in increasing mass. In each trial, we compare the simulated mass spectrum $M(i)$ with those generated from Eq.\ (\ref{equ:mf}) across a grid of \mlo\ and \mhi, determining the quantity $\chi^2$ at each point in the grid. The best fit values of both \mlo\ and \mhi\ are determined from the location in the grid of the $\chi^2$ minimum ($\chi^2_{\rm min}$), and the 68\% confidence limits on each parameter from the region of the parameter space defined by $\chi^2 = \chi^2_{\rm min} + 2.3$ \citep[following ][]{Avni76}.  We then repeat the experiment for each of the 1000 MC trials. As in K20, our results (see \fig{fig:kochMC}) show a similar cloud of points, centred close to the input values \mlo$_{\rm, in}$ and \mhi$_{\rm, in}$ but with a small systematic offset, and with points distributed about the median indicative of the random errors on \mhi\ and \mlo. 


\begin{figure}
\begin{center}
\includegraphics[width=8.5cm]{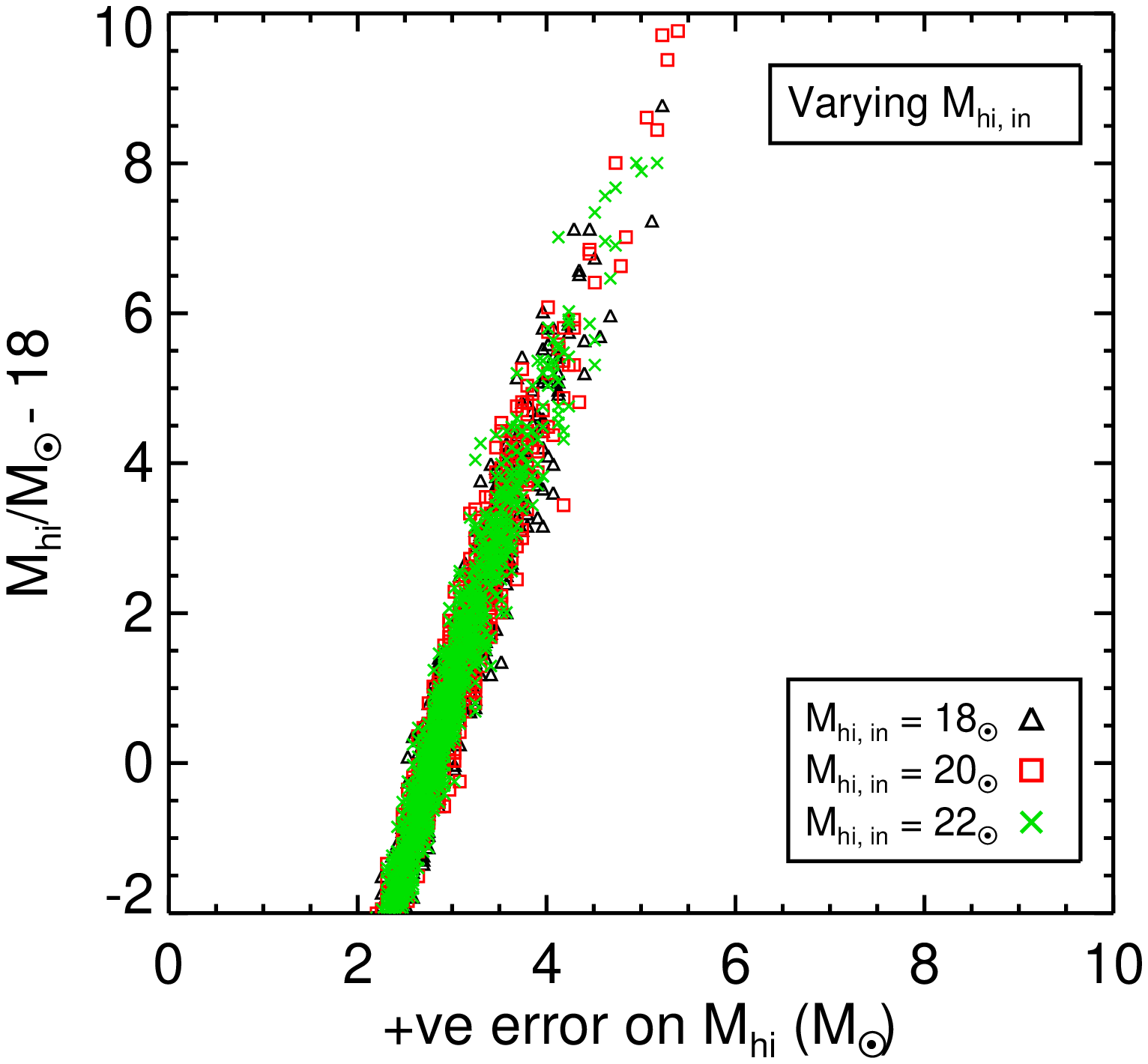}
\caption{Our version of K20's Fig.\ 5, showing the quantity (\mhi\ - 18\msun) for each of our MC trials, as a function of the positive error bar on \mhi\ for that trial. In addition to showing the results for the same input value as K20 (\mhi$_{\rm, in}$=18\msun), we also show the results for \mhi$_{\rm, in}$=20\msun\ and 22\msun. The exact same trend is seen irrespective of \mhi$_{\rm, in}$, demonstrating that the correlation seen in the plot {does not depend on the input value of \mhi}. }
\label{fig:kochmhi}
\end{center}
\end{figure}

\begin{figure}
\begin{center}
\includegraphics[width=8.5cm]{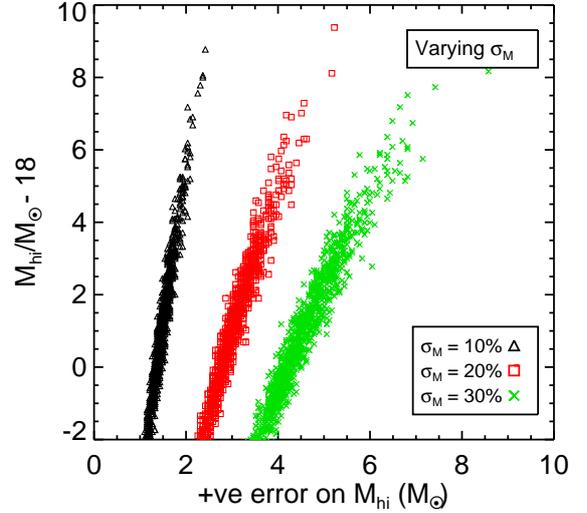}
\caption{Same as our \fig{fig:kochmhi}, but for fixed \mhi$_{\rm, in}$=18\msun, and with three different sizes of error bars on the input progenitor masses ($\sigma M$=10\%, 20\% and 30\%). Different trends are seen for each value of $\sigma M$, demonstrating that the correlation seen in K20's Fig.\ 5 is in fact caused by simple error propagation.}
\label{fig:kocherror}
\end{center}
\end{figure}

\subsection{The misinterpretation of the mass-error correlation}

\newnewtext{The next step taken by K20 is to take the upper error bars on each MC trial's measured \mhi\ (i.e. the upper error bars on the data-points in their Fig.\ 3, or our \fig{fig:kochMC}), and plot them as a function of $(M_{\rm hi} - M_{\rm hi, in})$ (shown in their Fig.\ 5). K20 name this latter quantity `overestimate of \mhi', but which is equivalent to `\mhi/\msun\ -18', i.e.\ the difference between the output and input values of \mhi. K20 observed that these two quantities are correlated, and interpreted this correlation as each randomly-scattered data-point being aware of how far it is from the centroid of the input distribution, and that \newtext{this information is communicated} back to the observer through the data-point's error bar. Continuing with this line of reasoning, K20 then claim that it is possible to transpose the results of each individual MC trial back to the input value of \mhi$_{\rm, in}$ with a precision of $\pm0.8$\msun. Finally, K20 then treats the analysis of the real-world sample of 24 progenitor mass estimates as though it were a single MC trial, and concludes that the best-fit of \mhi$=19^{+4}_{-2}$\msun\ should be adjusted down to \mhi$=15.8\pm0.8$\msun. This reduction of \mhi, as well as its upper error bar, implies that the difference between \mhi\ and the mass associated with the H-D limit is now in excess of 10$\sigma$, leading K20 to state that ``the RSG Problem remains''.

In Figs.\ \ref{fig:kochmhi} and \ref{fig:kocherror} we use the results of our MC experiment to make similar figures to Fig.\ 5 in K20, again seeing a correlation. However, in \fig{fig:kochmhi}, we make the same plot but for three different values of \mhi$_{\rm, in}$. Were the individual data-points aware of of the true input value of \mhi$_{\rm, in}$, the three sets of data would be offset in the $y$-direction, since a different correction factor would be required for each different value of \mhi$_{\rm, in}$. However, the exact the same trend is seen {\it irrespective of \mhi$_{\rm, in}$}. Hence, we can say that the trend seen in Fig.\ 5 of K20 has no dependence on \mhi$_{\rm, in}$. 

Next, in \fig{fig:kocherror} we again recreate Fig.\ 5 of K20, but for MC experiments with fixed \mhi$_{\rm, in}$=18\msun\ and three different values of $\sigma_M$ (10\%, 20\% and 30\%) which are typical of those in the current sample of progenitor masses (see DB18). In each case we see a correlation, but this time the slopes and offsets of these trends are dependent on $\sigma_M$. Our results show that the trend observed in Fig 5 of K20, rather than containing information on \mhi$_{\rm, in}$, is actually an illustration of how the random experimental errors on the progenitor masses propagate through to the error on \mhi.

We therefore conclude that K20 have misinterpreted the correlation seen in their Fig.\ 5. The scatter of the data-points in the MC trials about  \mlo$_{\rm, in}$ and  \mhi$_{\rm, in}$ (K20's Fig.\ 3 and our \fig{fig:kochMC}) illustrate the random experimental errors, and are driven by the uncertainties on the individual progenitor mass estimates. The errors on each progenitor mass are propagated from those on the host galaxy distance, the foreground reddening, and the bolometric correction, and represent the limits of our capability to measure these quantities. These cannot be corrected for. Instead, we assert that the trend seen in K20's Fig.\ 5, rather than being a means to correct for random errors, is in fact caused by the propagation of the observational errors into the uncertainty on the inferred value of \mhi. Using this trend to adjust the best-fit \mhi\ would be erroneous, and would result in values of \mhi\ that were systematically low.}

\subsection{The cumulative mass distribution}
Further evidence that the upper mass cutoff quoted by K20 is not supported by the data is found by a comparison to the numbers being fitted in that paper. In their analysis, K20 took the same photometric data, galaxy distances, reddenings and bolometric corrections as in DB18, in order to prove that there were issues with the latter paper's analysis. Hence, the masses of the individual progenitors being analysed are the same in both DB18 and K20. We can therefore assess the quality of the fits in both papers by overplotting the implied cumulative mass spectrum using the measured \mlo\ and \mhi\ and Eq.\ (\ref{equ:mf}). 

\begin{figure}
\begin{center}
\includegraphics[width=8.5cm]{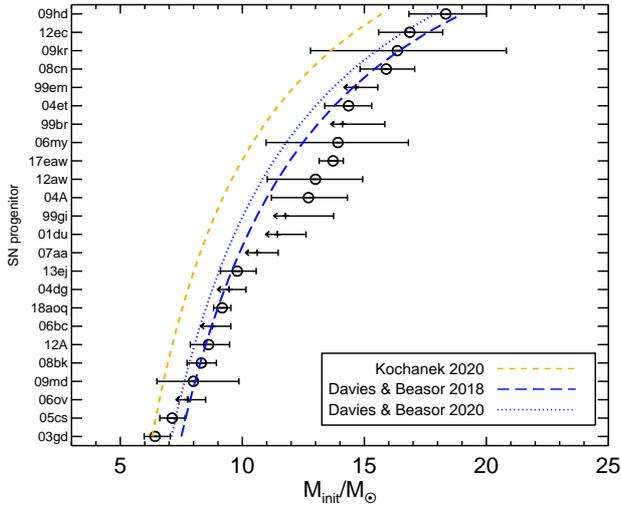}
\caption{The cumulative progenitor mass distribution analysed by both DB18 and K20. The blue dashed line shows the best-fit of DB18; the orange short-dashed line shows the prediction if \mhi=15.8\msun\ by K20 is adopted. For completeness, we show the fit of DB20 converted to the mass plane. }
\label{fig:fits}
\end{center}
\end{figure}

The fits of DB18, the more sophisticated analysis of DB20 (when converted from the luminosity plane to the mass plane), and that of K20, are overplotted on the progenitor mass distribution from DB18 in \fig{fig:fits}. It is clear from this figure that the K20 mass limits are inconsistent with the data. The predicted mass distribution is systematically offset to low masses for all progenitors, and in all but three progenitors this offset is greater than the quoted 68\% uncertainties. To quantify the quality of the K20 fit, we approximate the probability distributions on each detection as being asymmetric gaussians, and determine the root mean square of the quantity $z = (M_i-M_{K20})/\sigma_i$, i.e. the average deviation from the data in units of $\sigma_i$. We then integrate a normal distribution between $-\infty$ and $z$. We find that the K20 estimate of \mhi\ can be ruled out at the 99.6\% confidence level. 

\section{The RSG-BH production rate} \label{sec:bh}
One of the most exciting prospects offered by a statistically significant RSG Problem is the possibility that we could see a RSG disappear and form a BH in real time. Several research teams are currently monitoring nearby star-forming galaxies in the hope of catching such an event. To date, only two such candidates exist, neither conclusive \citep{Reynolds15,Adams17}. 

As time goes on, and no vanishing RSGs are confirmed, it is worth asking the question at what point does the {\it lack} of a detection become interesting in its own right? And, can this lack of any detection be used to make an independent estimate of \mhi? To answer these questions, we perform a series of simple MC experiments. First, we take the observed IIP rate in the nearby Universe\footnote{\newtext{Here, `nearby Universe' means close enough to be able to resolve and detect individual RSGs down to the mass threshold for BH production, which at the present time is $\sim$30Mpc.}} as being the number of events in the DB20 sample divided by the time since the first event. This yields a IIP rate of 1.3yr$^{-1}$. Next, we assume that stars may die as RSGs with initial masses up to 25\msun\ (see earlier), but that only stars with masses between 7\msun\ and \mhi\ will produce IIP SNe. Under the assumption of a Salpeter IMF, we can then estimate the RSG-to-BH rate as a function of \mhi. Next, randomly sampling from a Salpeter power law between \mhi\ and 25\msun\ at the appropriate BH formation rate, we run a series of MC experiments to determine the most likely number of RSG-to-BH events observed over a given time window. From the number of MC trials in which no RSG-to-BH event was observed, we can determine the probability that we would observe {\it no} such events as a function of observing time and of \mhi. 

\begin{figure}
\begin{center}
\includegraphics[width=8.5cm]{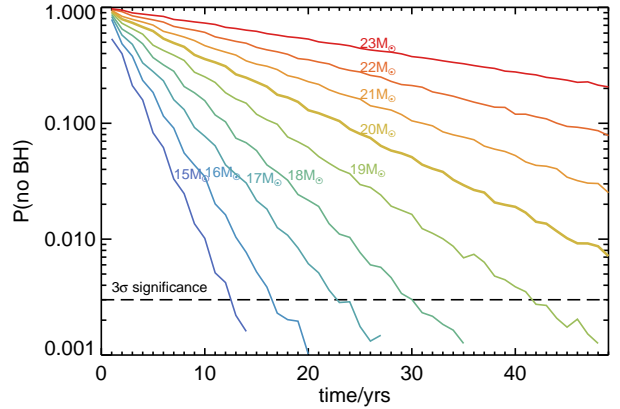}
\caption{The probability of observing no RSG collapse directly to a BH as a function of observation time, for a range of upper mass cutoffs \mhi. The 99.7\% confidence limit, analagous to 3$\sigma$, is indicated by the dashed line. }
\label{fig:BH}
\end{center}
\end{figure}

The results of this test are plotted in \fig{fig:BH}\footnote{\newtext{The plot may be transformed from the initial mass $M_{\rm init}$-plane to the terminal luminosity $L_{\rm fin}$-plane by applying the calibration $\log(L_{\rm fin}/L_\odot) = A + B\log(M_{\rm init}/M_\odot)$, where the constants $(A,B)$ are (2.67,2.02) and (2.92, 1.82) for the STARS and rotating Geneva models respectively. }}. Specifically, we plot the probability of finding zero RSG-to-BH events within a given time period for several different values of \mhi. Also indicated in the plot is the 0.3\% probability threshold, below which the significance of no observed RSG-to-BH event is greater than 3$\sigma$. For a survey which had been running for $\sim$12 years \citep[e.g.][]{Kochanek08}, we see that the lack of any detection argues against a value of \mhi\ below 15\msun. Within 5 years, K20's estimate of 15.8\msun\ could also be excluded. However, to provide a stringent independent test of the \mhi\ inferred from archival pre-explosion imaging on a timescale shorter than $\sim$decades, the search volume would have to be dramatically increased, with e.g.\ JWST.

\section{Conclusion -- a consensus on \mhi} \label{sec:conc}
\newnewtext{In this letter we have argued that the conclusion of \citet{Kochanek20} that ``the Red Supergiant Problem remains'' is invalidated by a statistical misintepretation in that paper. Specifically, a correlation between the best-fit upper mass cutoff \mhi\ and its error bar observed in Monte-Carlo tests was misconstrued as a means to correct for random experimental errors. Without this correction,} Kochanek's analysis of the IIP progenitor mass distribution finds \mhi$=19^{+4}_{-2}$\msun\ (using the STARS \minit-\lfin\ relation, MLR) is in excellent agreement with that of \citet{Davies-Beasor20} (\mhi$=18^{+4}_{-2}$\msun, STARS MLR), not just in terms of the best-fit value but also in terms of the probability distribution. With the same result being obtained seemingly independent of analysis strategy, we take this as evidence that the field is reaching consensus as to (a) the most likely value of \mhi, and (b) the precision to which this value can be quoted given the available data. Given this state-of-affairs, we can only reiterate our conclusions from \citet{Davies-Beasor20} that any mismatch between the luminosity distributions of field RSGs and of IIP progenitors cannot be established beyond the 3$\sigma$ level without at least a doubling of the sample size of progenitors.

\section*{Acknowledgements}
We thank the two anonymous referees and the editor for comments and suggestions which helped us improve our paper. We also thank Nate Bastian and Phil James for comments on the draft, and Chris Kochanek for on-going discussions. This work has made use of the IDL astronomy library, available at {\tt https://idlastro.gsfc.nasa.gov}. EB is supported by NASA through Hubble Fellowship grant HST-HF2-51428 awarded by the Space Telescope Science Institute, which is operated by the Association of Universities for Research in Astronomy, Inc., for NASA, under contract NAS5-26555.




\bibliographystyle{mnras}
\bibliography{/Users/astbdavi/Google_Drive/drafts/biblio} 



\bsp	
\label{lastpage}
\end{document}